\documentclass[letterpaper, 10 pt, conference]{ieeeconf}  % Comment this line out if you need a4paper

\IEEEoverridecommandlockouts                              % This command is only needed if 
\usepackage{graphics} % for pdf, bitmapped graphics files
\usepackage{epsfig} % for postscript graphics files
\usepackage{mathptmx} % assumes new font selection scheme installed
\usepackage{times} % assumes new font selection scheme installed
\usepackage{amsmath} % assumes amsmath package installed
\usepackage{amssymb}  % assumes amsmath package installed
\usepackage{hyperref}
\usepackage{multirow}
\usepackage{url}
\usepackage{array}
\usepackage{makecell}
\usepackage{graphicx}

\title{\LARGE \bf
User Based Design and Evaluation Pipeline for Indoor Airships
}

\author{Zhaoliang Zheng$^{1}$, Jiahao Li$^{1}$, Parth Agrawal$^{1}$ \\Zhao Lei$^{1}$, Aaron John-Sabu$^{1}$ and Ankur Mehta$^{1}$ % <-this % stops a space
\thanks{$^{1}$Authors are with the Department of Electrical and Computer Engineering, University of California Los Angeles, Los Angeles, California, 90095, USA
        {\tt\small \{zhz03, ljhnick, parthagrawal24, zlei, aaronjs, mehtank\}@ucla.edu}}%
}

\begin{document}

\maketitle
\thispagestyle{empty}
\pagestyle{empty}

%%%%%%%%%%%%%%%%%%%%%%%%%%%%%%%%%%%%%%%%%%%%%%%%%%%%%%%%%%%%%%%%%%%%%%%%%%%%%%%%
\begin{abstract}

Designing a controllable airship for non-expert users or preemptively evaluating the performance of desired airships has always been a very challenging problem. This paper explores the blimp design parameter space from the aspect of the user by considering various distributions of thrust, combinations of propulsive mechanisms, and balloon shapes. We provide open-source modular hardware and reconfigurable software design tools that allow inexperienced users to design a custom airship in a short time. Based on these design parameters, this paper develops a more engineering-focused evaluation system that can characterize the performance of different indoor blimps. An analytical comparison and some case studies that consider various points in the design parameter space have been conducted to prove the feasibility and validity of our design and evaluation system.

\end{abstract}

%%%%%%%%%%%%%%%%%%%%%%%%%%%%%%%%%%%%%%%%%%%%%%%%%%%%%%%%%%%%%%%%%%%%%%%%%%%%%%%%
\section{INTRODUCTION}
  Due to their ability to neutrally float in the air, lighter-than-air vehicles (LTAVs) had been widely studied and used as research and military platforms for aerial robotics over the last century. As a branch of LTAVs, the indoor airship is gaining increasing attention due to its promising potential for many applications \cite{sebbane2011lighter}. In the past 20 years, indoor blimps have been developed for infrastructure inspection \cite{nitta2017visual}, environmental data collection \cite{kantor2001collection}, indoor localization and mapping \cite{muller2013efficient}, education and research platforms \cite{gorjup2020low}, vision-based human-robot interaction \cite{NYao2017}, and other activities. While these tasks may also be conducted using unmanned aerial vehicles (UAVs) such as quadrotors, their flight duration time is generally only between 20 to 30 minutes, restricted by the power required for them to hover in the air. Quadrotors can also cause safety concerns when they operate in indoor environments with humans due to their relatively high operating speed and high-speed rotating blades.

\begin{figure}
    \centering
    \includegraphics[width=0.48\textwidth]{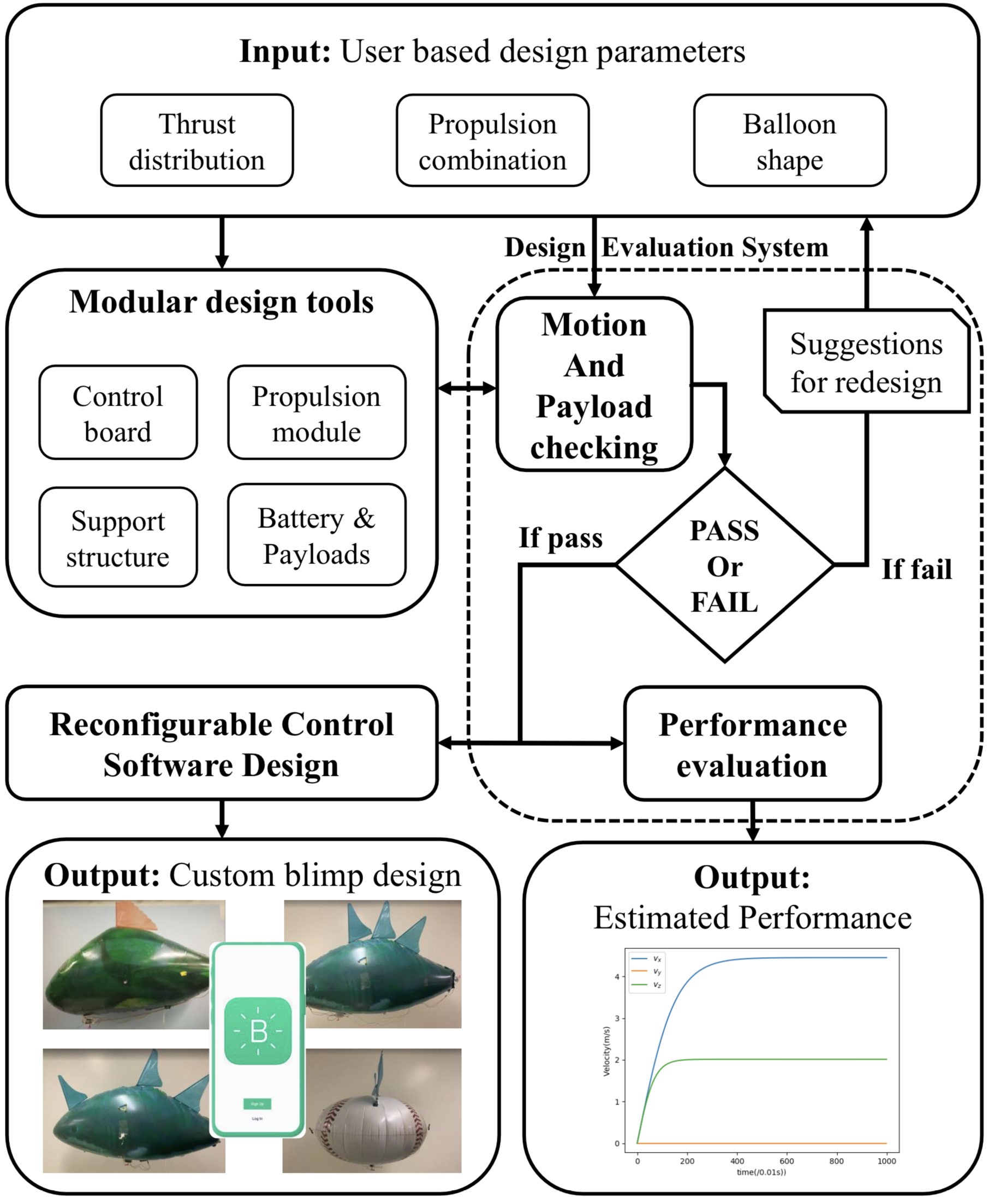}
    \caption{We provide a user-based blimp design pipeline that allows inexperienced users to design and evaluate custom blimp designs. The modular design tools allow users to customize their hardware and electronics, and the reconfigurable control software allows users to change their control method under different configurations from the software level. The design evaluation system can help users check whether their design meets the motion and payload criteria, in addition to outputting an estimated performance of their designs.}
    % a modular design tool that could allow user to design their own blimps using the high-level parameters considering different thrust distribution, propulsion combination, different balloon shapes and weight distribution. And based on those user design parameters, our evaluation system could estimate the performance of that particular design.
    \label{fig:1}
\end{figure}

In recent years, several research projects have focused on developing techniques related to indoor airships, and oftentimes, such projects focused on the controller design for a particular blimp developed by the authors. This has been observed with respect to PID altitude control \cite{wang2017altitude}, fuzzy logic control \cite{gonzalez2009developing}, neural network control \cite{zufferey2006flying} and other controllers with various applications \cite{takaya2006motion}\cite{kawamura2008learning}. These controllers only apply to their specific scenarios and particular airships, however, and are not accessible to ordinary users interested in implementing a novel design in user-specific scenarios such as a big warehouse or supermarket.  

% As for the control of indoor airships, Wang et al. designed a PID altitude controller for their oval shape indoor airships\cite{wang2017altitude}. PID and fuzzy logic controllers are designed and composed to ensure the safety of designed autonomous blimp in the work \cite{gonzalez2009developing}. A neural network control architecture and input processing method for indoor oval shape blimps is presented in the work\cite{zufferey2006flying}. PID controllers are design to conduct motion control and landing in the work \cite{takaya2006motion} \cite{kawamura2008learning} for their cylinder indoor blimp.  The work \cite{burri2013design} presented a spherical blimp design and corresponding control algorithm to control its motion. The authors of the work \cite{cho2017autopilot} developed an autopilot system that considered speed, heading and altitude into their controller and implemented that autopilot design on their miniature saucer-shape blimp. There papers have detailedly discussed how to design controllers for the speed, heading and altitude of indoor blimps. However, their work didn't cover too much about the blimp design question but briefly mentioned their own particular design.  

Among papers that deal with blimp design and application, 
%an efficient physics-based dynamic modelling of indoor airships including a pragmatic methodology for parameter identification without the need for complex or costly test facilities was presented by Zufferey et al \cite{zufferey2006flying}. 
%In the work \cite{biju2017design}, 
Biju et al. described their blimp design subsystem and conceptual design method. Their blimp was very large in size, however, and difficult to scale to a miniature blimp \cite{biju2017design}. In contrast, \cite{gorjup2020low} presented a low-cost open-source miniature indoor airship for research and education applications, but its fixed structure makes it hard to extend its capability.
% controller board Raspberry Pi Zero will consume lots of power and given the battery they have on board, it's operating time is not optimistic.
The authors of papers \cite{cho2017autopilot} and \cite{tao2018parameter} presented GT-MAB, a class of indoor miniature blimp designs, and a method to identify its model parameters. This approach is only applicable to a saucer-shaped balloon, however, and the fixed gondola complicates extending the concept to other configurations.      

The research presented in this paper focuses on the design and evaluation aspect of indoor airships as shown in Fig. \ref{fig:1}. We propose a modular design approach that allows non-expert users to build and design their own custom indoor blimps. The modular design tools are open-source and can be easily fabricated using 3D printers. To help users have a better understanding of designing custom indoor blimps, our work explores the blimp design parameter space from the user perspective by considering different configurations that users may care about, i.e. different thrust distributions, different propulsion combinations, varied balloon shapes, and the weight distribution. 

%The thrust distribution and be configured in a way that the user wants, but the thrust distribution has to satisfy the basic motion primitives, which our design and evaluation system will automatically check for users. In propulsion combination parameter space, it contains different "motor + propellers" combinations, and here in our paper, we provide 7 different combinations. Considering the feasibility of manufacturing balloon envelopes, three different common shapes of balloons are considered in this paper, which are sphere shape, saucer shape, and oval fish shape. As for the weight distribution, we analyzed the impact of decentralized and centralized weight distribution in this paper. 

%Based on the , we derived the low-level parameters that can be feed into our design and evaluation system. 
Using user-based design parameters as inputs, our evaluation system then checks that these configurations can successfully generate basic motion primitives while estimating the payload capacity to ensure lighter-than-air travel. If the design from users passes these motion and payload checks, the evaluation system will output the blimp performance, defined as the maximum horizontal and vertical velocity the blimp can achieve (for a detailed explanation, see section II). 

% We expect our evaluation model will eventually generate an estimated performance of the designed blimp which includes the satisfaction of motion primitives, payload specification, and maximum horizontal and vertical velocity.     

% Difficulties:\begin{itemize}
%     \item How to validate the evaluation system?
%     \item How to frame the design tool problem in the paper?
%     \item How to solve for the fin aerodynamic problem?
%     \item What kind of test should be done to verify that problem? 
% \end{itemize}

% Motivation:\begin{itemize}
%     \item Redistributed the configuration so that it could fit into 
% \end{itemize}

% As for blimp design, an efficient physics-based dynamic modelling of indoor airships including a pragmatic methodology for parameter identification without the need for complex or costly test facilities was presented by Zufferey et al\cite{zufferey2006flying}. M. Burri et al. presented a novel spherical omnidirectional blimp\cite{burri2013design}. Paper \cite{Tao2018} presented the identification of the rotation-related parameters of the blimp dynamics model through swing motion of the robot.  

% As for blimp control, Wang et al. presented an altitude controller for an indoor blimp\cite{wang2017altitude}.Hiroaki Fukushima et al presented a model predictive control method of an autonomous blimp\cite{Hiroaki2006}.   

% As for blimp perception, Ningshi Yao et al. presented an approach that utilized monocular camera to allow blimp to detect and follow human\cite{NYao2017}.

In summation, the main contributions of our paper are the following: 

% 1. exploring the user-based design parameter space considering various configurations in thrust distribution, propulsion combination, balloon shape, and weight distribution,

1. We create a modular design tool that allows users to design their own custom blimp, along with a software configurable motion control panel that can correctly interpret user-defined hardware electronics connections, such as motors to a controller board.

2. We propose a custom-blimp-design evaluation system that takes in design parameters, checks configuration feasibility, then provides as output the steady-state performance of the blimp design.

% 4. Explore the blimp design space

% 5. Provide the reward function for later reinforcement learning-based blimp design generator

% 6. Validate the performance evaluator by building a blimp and comparing the actual performance to the predicted performance

\section{User based design parameters space}

\begin{figure*}[]
    \centering
    \includegraphics[width=0.99\textwidth]{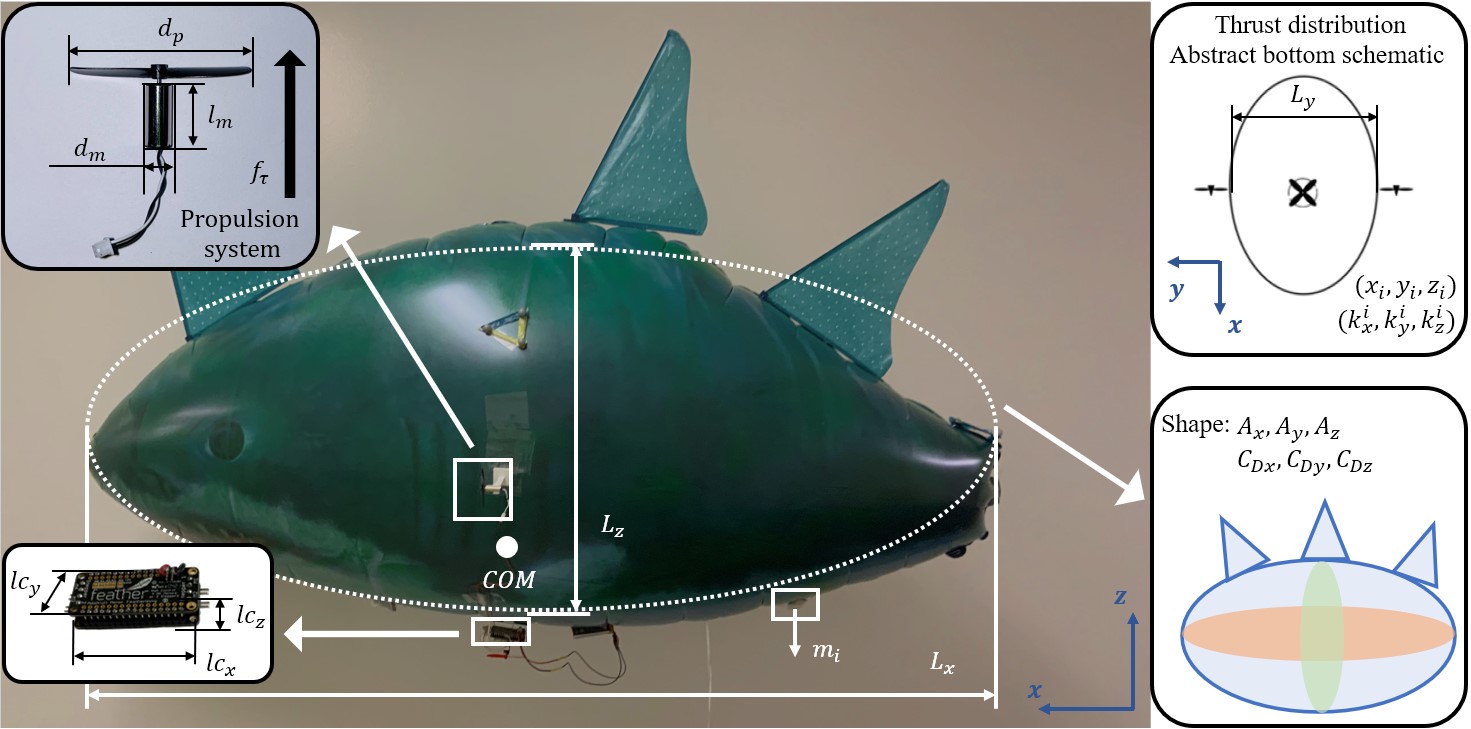}
    \caption{The user-based parameters should mainly consider thrust distribution, propulsion system and the balloon shape}
    \label{fig:param_space}
\end{figure*}

Basically, for normal users who would like to customize their own airship for some particular scenarios, there are three aspects that are required to consider in their designs: thrust distribution, propulsion system, and balloon shape. Our design parameters are also derived from these three aspects.

\subsection{Thrust distribution}
In many indoor blimp designs, different thrust distribution is the main difference between the different designs. In thrust distribution, what users should care about is the position of their thrust. In order to enable the airship to have complete space exploration capability, the thrust distribution should be designed to satisfy three motion primitives: 
\begin{itemize}
    \item Maintaining forward speed. The blimp should be able to maintain a desired constant forward speed while having zero vertical speed and zero yaw angular speed. \item Changing altitude:  The blimp should be able to ascend or descend to the desired height. 
    \item Changing orientation: The blimp should be able to spin in place so that its yaw angle can be stabilized at any desired value.
\end{itemize}

\subsection{Propulsion system combination}
The propulsion system that we talk about in this paper is limited to only the "DC motor + propeller + servo motors(optional)" combination. Motors propellers have been studied for more than centuries. And they are dynamically stable, reliable, easy to design and control \cite{glauert1935airplane}.

One benefit of using DC brushed motors is that the relationship between motor voltage $V_{motor}$ and thrust $\tau$ generated by propellers is a fixed function as shown below:
\begin{equation}
    V_{motor} \xrightarrow[propeller]{load} \tau
\end{equation}
By changing the PWM signals from the control board, we can control the voltage outputs of the motor and thus control its thrust. 

\subsection{Balloon shape}
 
In this paper, we only discuss the balloon envelope made of Mylar material, which is a polyester film made from stretched polyethylene terephthalate (PET). This material is chosen because it is used for its high tensile strength, chemical and dimensional stability, transparency, reflectivity, gas and aroma barrier properties, and electrical insulation \cite{mylar}. Based on most of the literature, in our paper, we only discuss three common shapes: sphere, saucer, and oval shape balloon. Different shapes of balloons have different 3D dimensions and aerodynamic properties which will affect the performance of the designed airship.

\subsection{Design parameters}

In the design parameters space, we considered two types of design parameters that are user-friendly: hardware design parameters that are used for designing gondolas and support materials, evaluation design parameters that can be fed as inputs into the evaluation system.

For hardware design parameters, we need to consider:\begin{itemize}
    \item Propeller diameter: $d_p$
    \item Motor length: $l_m$
    \item Motor diameter: $d_m$
    \item Dimension of the selected control board: $lc_x,lc_y,lc_z$
\end{itemize}

For evaluation design parameters, we need to consider:\begin{itemize}
    \item Thrust of propeller: $f_\tau$\
    \item The position of each thrust component $i$ in the body frame: $p_i:=[x_i,y_i,z_i]^T$ s.t. $i \in \{1...N\}$, N is the total number of thrust vectors in the system.
    \item The orientation primitives of each thrust component $i$: $K_i:[k_x,k_y,k_z]^T$ s.t. $k_x,k_y,k_z \in \{0,1\}$, 1 means that this thrust vector is in line with the corresponding axis and 0 means that it's not generating thrust on that axis.
    \item The cross sectional area (CSA) of the inflated balloon: $A_xy,A_yz,A_xz$.
    \item Mass of electronics, balloon envelope, support materials, and payload:  $m_{elec},m_{envelop},m_{sup},m_{payload}$
\end{itemize}

Look at the design of indoor blimps in recent years, miniature indoor blimps actually have a relatively simple design framework that contains several necessary components. The fundamental difference between different designs is the configuration of thrust, propulsion system, balloon shape, and weight distribution of sensors and other components. 

\subsection{Modular hardware and reconfigurable design}
\begin{figure}
    \centering
    \includegraphics[width=0.48\textwidth]{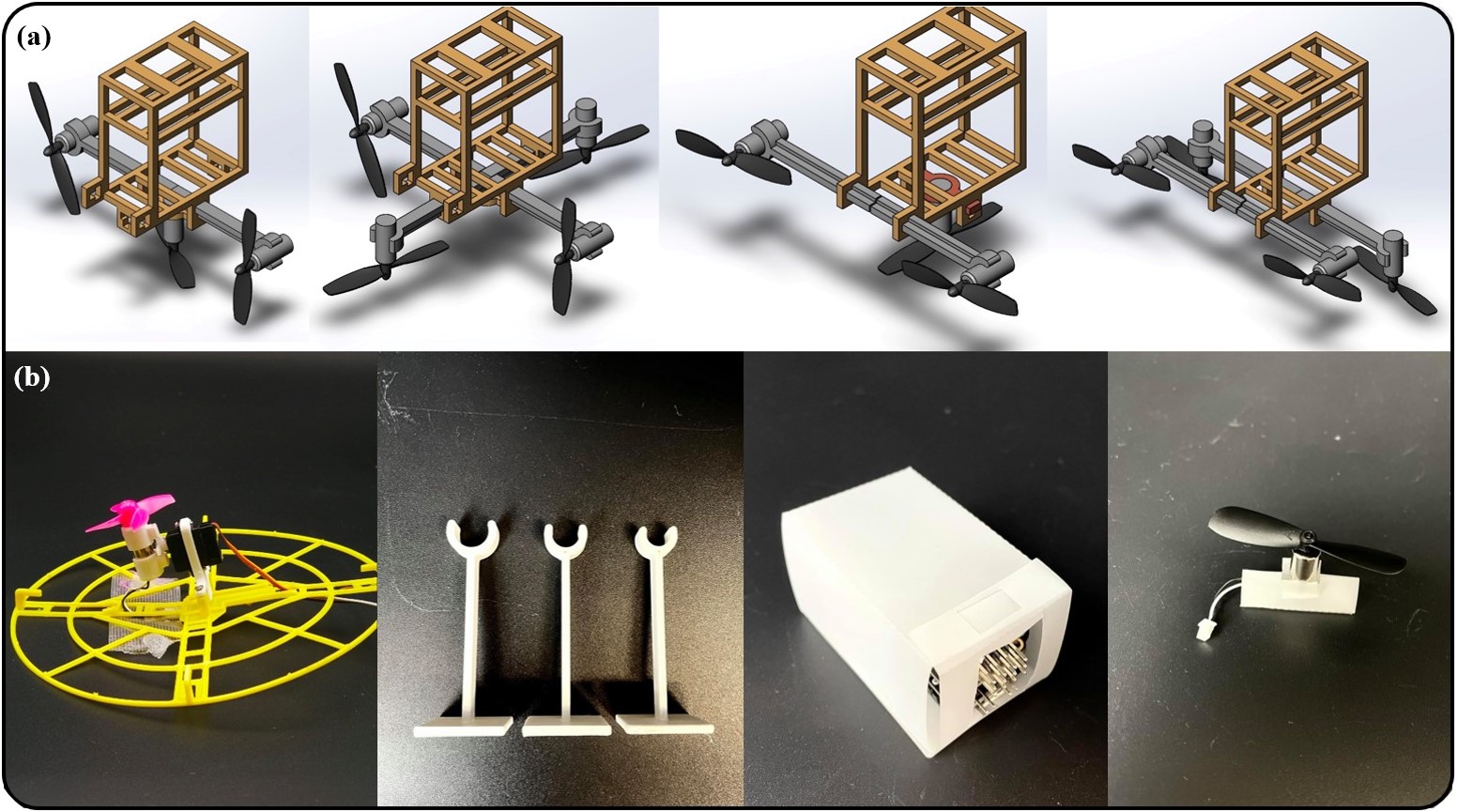}
    \caption{(a) centralized reconfigurable gondola design. (b) decentralized reconfigurable gondola and motor support design.}
    \label{fig:3}
\end{figure}
The purpose of the modular re-configurable hardware and design tool is (1) to help users to allocate their own thrust distribution (2) to provide suitable support materials that fit the dimension of the chosen propulsion system.

We designed a reconfigurable 3D printed gondola which allows users to reconfigure their thrust distribution in the way they want by simply changing the position of the motor support as shown in Fig. \ref{fig:3}(a). For decentralized design, we also provided separated support structures that allow users to place their control board, DC motors, servo motors, and propellers on any position of the balloon surface as shown in Fig. \ref{fig:3}(b). Based on different design purposes and configurations, different types of gondolas and support materials can be easily fabricated using 3D printers or origami printers. Based on the user's required  hardware design parameters that were introduced before, all these support materials can be customized using our open source files provided in the git: \url{https://github.com/zhz03/User_Based_Design_and_Evaluation_Pipeline}

As for different dimensions of the propulsion system, our support materials could customize the diameter of motors between 4mm $\sim$ 8.5mm. The shaft diameter of the propeller can support common types such as 1 mm and 0.8 mm on the market and it's determined by the shaft diameter of the selected motors. 

\subsection{Reconfigurable control software design}

In the blimp design, reconfigurable hardware is only one part of the design, the other part of it is to allow users to control their designed blimps in the right way. The purpose of reconfigurable software is to enable the operation of the control software to match the right wire connection between the control board and propulsion system after the modular hardware design meets basic motion primitives. In our design pipeline, we choose Adafruit ESP32-based Feather and its compatible 4-Channel DC Motor as well as 8-Channel PWM Servo FeatherWing as the main control circuit.

\begin{figure}[h]
    \centering
    \includegraphics[width=0.48\textwidth]{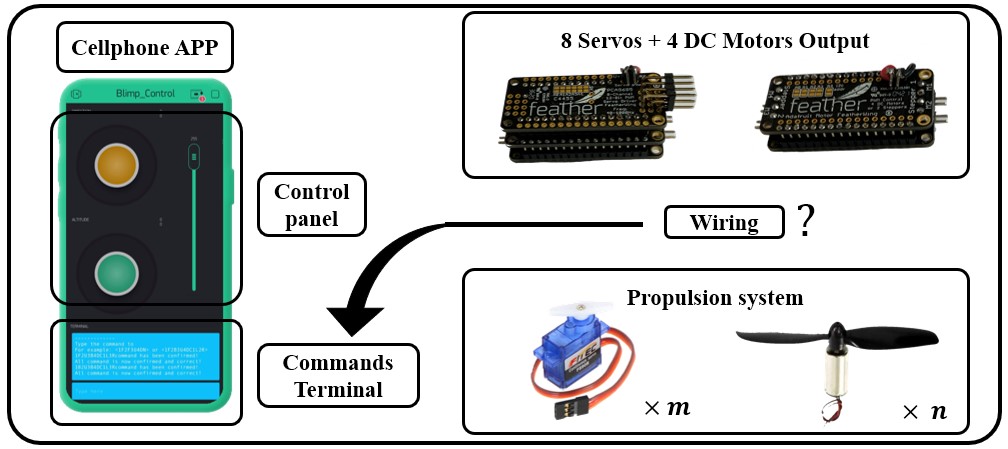}
    \caption{Reconfigurable software and its working principle}
    \label{fig:2}
\end{figure}

We choose Blynk APP as the control interface since it's a freely available application that anyone could download online.
As shown in the \ref{fig:2}, our control panel contains two joysticks to control blimp movement(horizontal and vertical) and one slider to control all motors speed. Since the wiring of electronics will directly determine how to manipulate the custom blimp, our reconfigurable low-level program can handle the wiring and connection between the control board and propulsion system and allows users to re-map the control to the actual propulsion system they design through the terminal panel.

To find out the correct command that matches the mapping, users need to go over the following three iterations:

    (Iter. 1) Initialization: Initial command "1F2B3U4DN" means that DC motor channel 1 is "forward" rotation, channel 2 is "backward" rotation, channel 3 is "upward" rotation, channel 4 is "downward" rotation, and the left and right direction is not confirmed. 
    
    (Iter. 2) Determine main horizontal and vertical channel: based on the actual propellers rotation and activation situation, use joystick to check if previous command is correct, if not, then change the command to determine the correct horizontal and vertical channel and rotation direction. Command after iteration example: "1F2F3U4DC1L2R". The last 5 digits means that horizontal and vertical channel are confirmed using DC motor and assume channel 1 to rotate left and channel 2 to rotate right. 
    
    (Iter. 3) Determine the rotation channel from previous command: if the previous command is correct then stop. If not, then switch the rotation channel, for example: "1F2F3U4DC2L1R". 

\section{Evaluation system design}

In our evaluation system, the problems we could like to help users to solve are: (1) Whether the designed blimp can effectively move in the 3D space, which is to meet the three basic motion primitives. (2) Whether the designed airship can be effectively suspended in the air, that is, it can maintain neutral buoyancy in the air without activating the propulsion system. (3) What is the estimated maximum performance of the designed airship. In this section, we will discuss three main functions in the evaluation system: (1) Motion primitives checking, (2) Payload checking and estimation, (3) Steady-state performance estimation. 

\subsection{Motion primitives checking}
When non-expert users use our modular design framework to design their own blimps. one notable design problem is that their custom design may not satisfy their needs of exploring the space, which is to satisfy the basic motion primitives as described before. Mathematically, motion primitives checking is defined as: 
\begin{equation}
    \mathbf{F}_p = \left[ \begin{array}{ccc}
F_{px} 	\neq 0\\
F_{py}\\
F_{pz} 	\neq 0
\end{array} 
\right],
    \mathbf{M}_p = \left[ \begin{array}{ccc}
M_{px} \\
M_{py}\\
M_{pz} \neq 0	
\end{array} 
\right]
\end{equation}
Where, $\mathbf{M}_p$ is the moments generated by propulsion force. $\mathbf{F}_p$ and $\mathbf{M}_p$ can be calculated by Equ. 
\begin{equation}
\label{equ:12}
    \mathbf{F}_p = \sum_{i} f_i \mathbf{K}_i
\end{equation}
\begin{equation}
\label{equ:13}
    \mathbf{M}_p = \sum_{i} \mathbf{a}_i \times (f_i \mathbf{K}_i)
\end{equation}

\subsection{Payload estimation and checking} 

\begin{table*}[h]
\label{table1:Design tool}
  \centering
    \caption{Two case study using modular hardware and reconfigurable software design tool}
  \begin{tabular}{ | c | c | c | c | m{1em} | c |  m{1em} | c |  m{1em} | c | }
    \hline
    \begin{tabular}[c]{@{}c@{}}Case\\ \end{tabular}
     & Circuit diagram & Thrust dist. & \begin{tabular}[c]{@{}c@{}}Motion\\Checking \end{tabular} & Iter. 1 &  Status & Iter. 2 &  Status & Iter. 3 &  Status  \\ \hline
    \begin{minipage}[b]{0.3\columnwidth}
		\centering
		\raisebox{-.5\height}{\includegraphics[width=\linewidth]{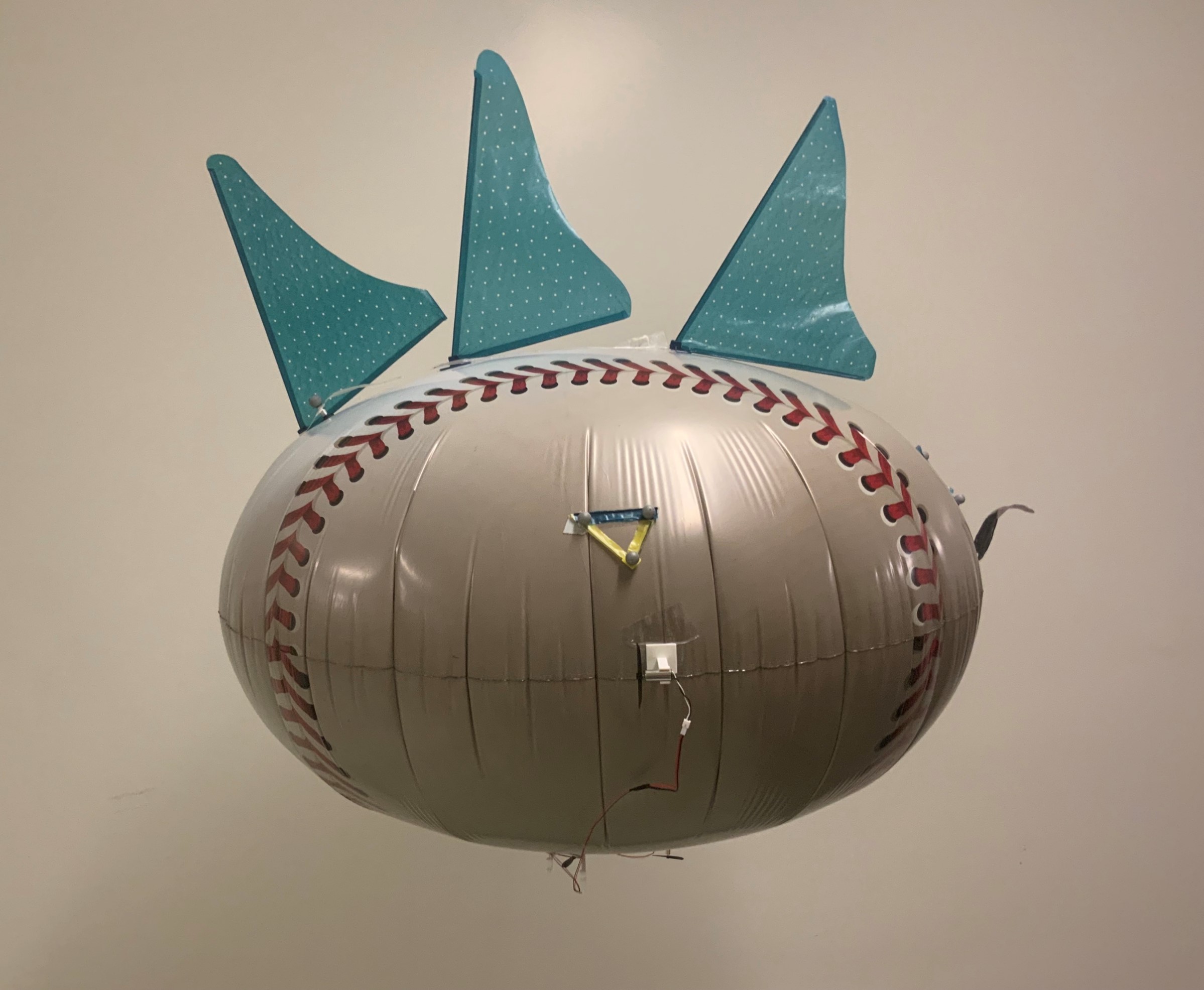}}
	\end{minipage}
    &     \begin{minipage}[b]{0.3\columnwidth}
		\centering
		\raisebox{-.5\height}{\includegraphics[width=\linewidth]{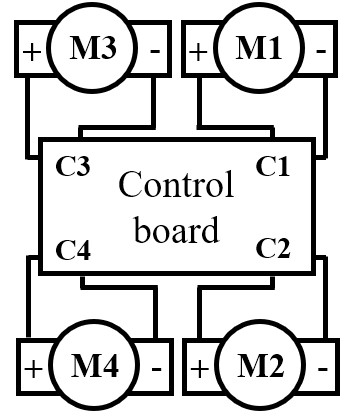}}
	\end{minipage}
    &  \begin{minipage}[b]{0.3\columnwidth}
		\centering
		\raisebox{-.5\height}{\includegraphics[width=\linewidth]{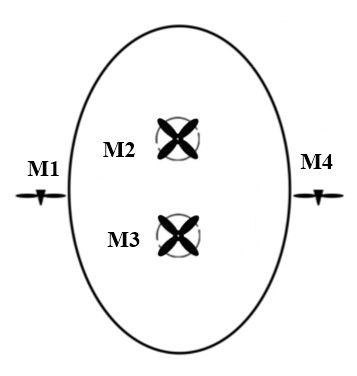}}
	\end{minipage}
	& Pass
    & \rotatebox{90}{"1F2B3U4DN"}
    & None
    & \rotatebox{90}{"1F2U3U4BC1L4R"}
    & \multirow{2}{*}{\begin{tabular}[l]{@{}l@{}}horizontal$\checkmark$\\ vertical$\checkmark$\end{tabular}} 
    & \rotatebox{90}{"1F2U3U4BC4L1R"}
    & \multirow{2}{*}{\begin{tabular}[l]{@{}l@{}}horizontal$\checkmark$\\ vertical$\checkmark$\\ rotation$\checkmark$\end{tabular}}
    \\ \hline
    \begin{minipage}[b]{0.3\columnwidth}
		\centering
		\raisebox{-.5\height}{\includegraphics[width=\linewidth]{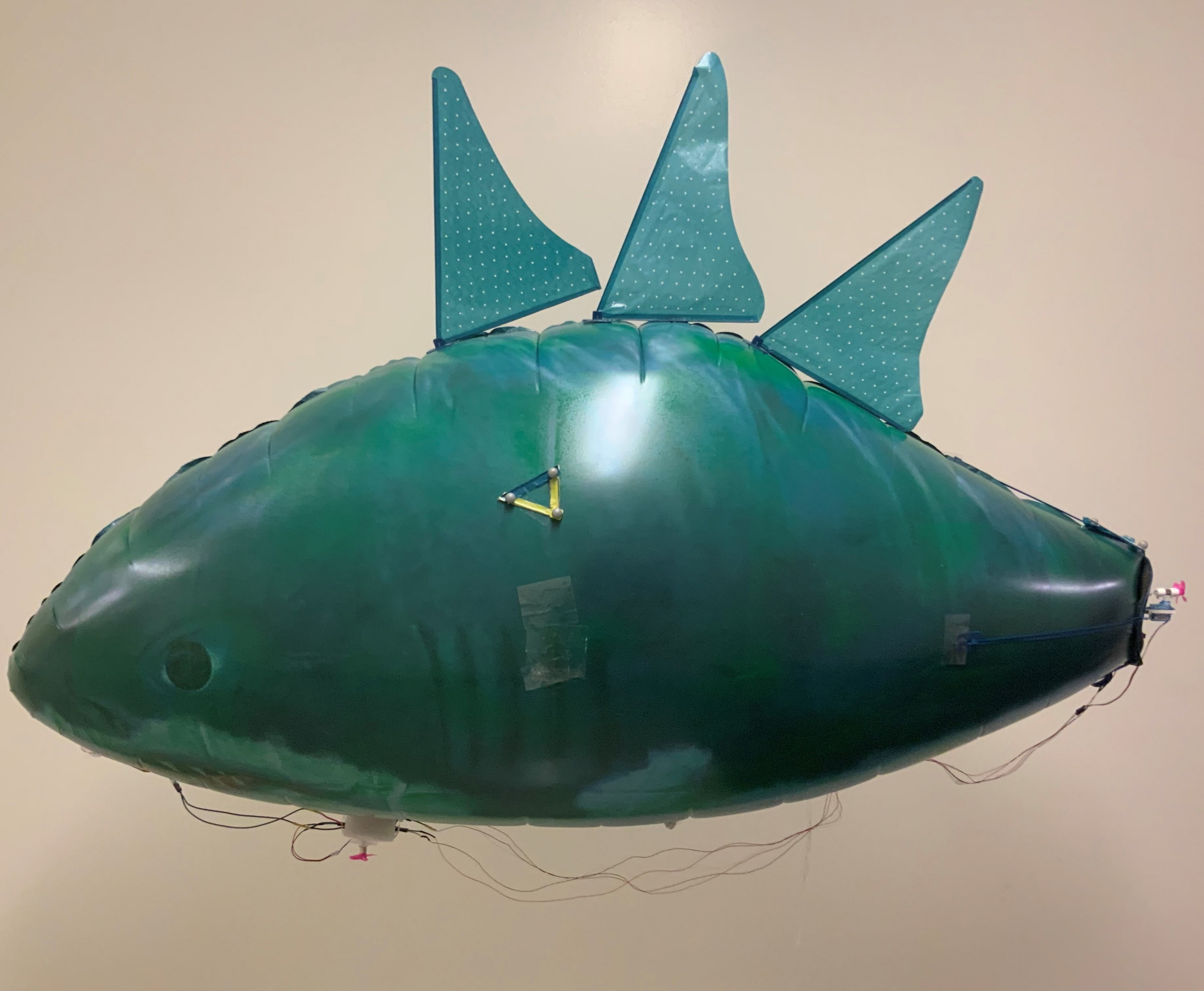}}
	\end{minipage}
    &     \begin{minipage}[b]{0.3\columnwidth}
		\centering
		\raisebox{-.5\height}{\includegraphics[width=\linewidth]{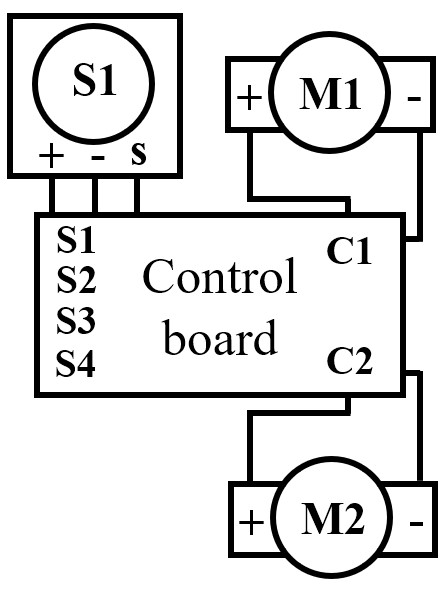}}
	\end{minipage}
    &  \begin{minipage}[b]{0.3\columnwidth}
		\centering
		\raisebox{-.5\height}{\includegraphics[width=\linewidth]{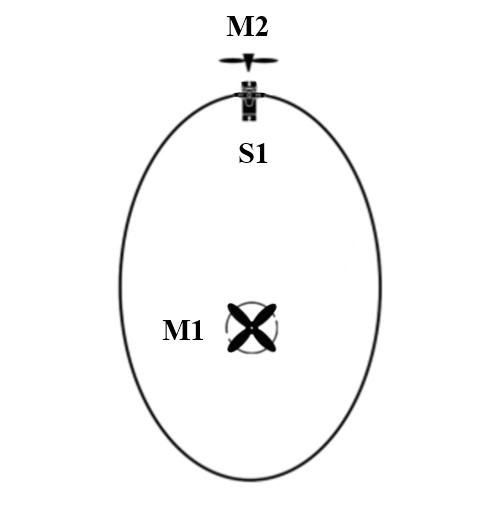}}
	\end{minipage}
    & Pass
    & \rotatebox{90}{"1F2B3U4DN"}
    & None
    & \rotatebox{90}{"1U2F3N4NS21MLR"}
    & \multirow{2}{*}{\begin{tabular}[l]{@{}l@{}}horizontal$\checkmark$\\ vertical$\checkmark$\end{tabular}} 
    & \rotatebox{90}{"1U2F3N4NS21MRL"}
    & \multirow{2}{*}{\begin{tabular}[l]{@{}l@{}}horizontal$\checkmark$\\ vertical$\checkmark$\\ rotation$\checkmark$\end{tabular}}
    \\ \hline
  \end{tabular}
\end{table*}

Oftentimes, for users who choose a 2D envelope to make an airship, it's hard to tell whether their chosen 2D balloon envelope could offer enough buoyancy after it's inflated. For indoor airship, the total lifting capacity can be divided into: electronics, balloon envelope, support materials and payload. Therefore, the total mess is calculated by:
    \begin{equation}
    \label{equ:6}
        m_{total} = \sum_{i} m_i = m_{elec} + m_{envelop} + m_{sup}+ m_{payload}
    \end{equation}  
    
To allow the designed airship to maintain neutral buoyancy in the air without activating the propulsion system. The payload checking criteria is: $ m_{payload} \geq 0$.

The buoyancy $F_B$ can be calculated using Archimedes' principle and is equal to the weight of helium and all other components:\begin{equation}
        F_B = \rho_{air} V_{envelop} g = \rho_{air} f(L_x,L_y,L_z) g = ( m_{heliem} + m_{other} ) g
        \label{equ:8}
    \end{equation}
    where $\rho_{air}$ is the density of air and $V_{envelop}$ is the volume of the envelope, which is the function of 3D geometry dimension $f(L_x,L_y,L_z)$. 

In this paper, the balloon shapes we discussed are sphere-shaped, saucer-shaped, oval-shaped, and non-regular oval-shaped. All of these four different shapes can be characterized and using the ellipsoid 3D geometry model to calculate their volume. 

According to Equ.(\ref{equ:6}) (\ref{equ:8}), the payload is calculated:
\begin{equation}
    m_{payload} 
    %& = & V_{envelop} (\rho_{air} -\rho_{helium}) - (m_{elec} + m_{envelop} + m_{support}) 
    % = \frac{4}{3} \pi  L_x L_y L_z (\rho_{air} -\rho_{helium}) - (m_{elec} + m_{envelop} + m_{sup})
    = V_{envelop} (\rho_{air} -\rho_{helium}) - (m_{elec} + m_{envelop} + m_{sup})
\end{equation}

However, $ L_x L_y L_z $ are the dimension after the balloon is inflated. To tell directly from 2D deflated envelop, we need to map the dimension of the 2D deflated balloon to $V_{envelope}$. 

Since the elasticity of Mylar is very small, by using the principle of entropy increase, we can assume that the surface area of 2D envelope $SA_{2D}$ is similar to the surface area of 3D inflated balloon $SA_{3D}$: $SA_{2D} \approx SA_{3D}$. Therefore, 
\begin{align}
    %SA_{2D}                      &= 4\pi a_{deflated} b_{deflated}\\
    %\text{But}, 
    \because SA_{3D} &= 4\pi r_{3D}^2 \approx SA_{2D}\\
    \implies r_{3D}              &= \sqrt{\frac{SA_{2D}}{4\pi}}\\
    V_{envelop}                 &= \frac{4\pi}{3} r_{3D}^3
\end{align}

\subsection{Steady-state performance estimation}

The kinematics and dynamics of our blimp system can be divided into a linear component and an angular component. We first show how we modeled the linear part, then we present a steady-state maximum speed calculation, which is a critical performance metric.
\subsubsection{Modeling}
The linear system dynamic in the world frame is: %\cite{GRASPTestBed}
\begin{equation}
m\begin{bmatrix}\dot v_x\\\dot v_y\\\dot v_z\end{bmatrix}=
\begin{bmatrix}0\\0\\mg-F_B\end{bmatrix}+
R_{wb}\begin{bmatrix}F_X-f_x(v_{x,\text{body}})\\F_Y-f_y(v_{y,\text{body}})\\F_Z+f_z(v_{z,\text{body}})\end{bmatrix}
\end{equation}
Where $m$ is the total mass of our blimp, $F_B$ is the buoyancy, $R_{wb}=R(\phi,\theta,\psi)$ is a rotation matrix that captures the controller-governed attitude with a roll $\phi$ pitch $\theta$ and yaw $\psi$, $F$ is the net propulsion provided by motors, and $f$ being the drag, which is a function of velocity in the body frame.

\subsubsection{Steady-state Solution}

We observe that our drag model is in fact a mapping from speed to drag force \cite{mueller2004development}. So if we can calculate what the terminal drag force is, we can infer what the terminal velocity should be. 

To calculate the terminal drag force, we set the acceleration on all 3 axes to be equal to 0, that is: $\dot v = 0$.

The terminal drag can thus be extracted from Equ. \ref{equ:ter}.
\begin{equation}
\label{equ:ter}
  \begin{bmatrix}f_x(v_{x,\text{max,body}})\\f_y(v_{y,\text{max,body}})\\-f_z(v_{z,\text{max,body}})\end{bmatrix}=
\begin{bmatrix}F_X\\F_Y\\F_Z\end{bmatrix}
-R_{bw}\begin{bmatrix}0\\0\\F_B-mg\end{bmatrix}  
\end{equation}

With our drag model, we can calculate the terminal velocity as in Equ. \ref{equ:ter_v}.
\begin{equation}
\label{equ:ter_v}
f=\frac{1}{2} \rho v^{2} C_{D} A
\implies v_\text{max,body}=\sqrt{\frac{2f}{\rho C_D A}}    
\end{equation}

Where, $C_D$ is the drag coefficient which can by calculated using the method provided in \cite{zufferey2006flying} or in our paper, we used CFD software -- Ansys Fluent to approximate. The Ansys fluent can calculate the result in half hours after the 3D model is designed and modified in our modular design.

\section{Case Study, Experiment and Analysis}

\subsection{Designed blimp case study}

To verify our design tools, we designed two different cases with different shapes and different numbers of motors and propellers. Both of them have passed our motion primitive checking since their thrust distributions meet three basic motion primitives requirements. Table I shows how to map the wiring with control software in a three-step iteration through commands. Eventually, after the adjustment through commands, the software can control the design airship to conduct the right horizontal, vertical, and rotation movement.  

\subsection{Balloon Volume estimation case study}

\begin{figure*} [h]
    \centering
    \includegraphics[width=0.95\textwidth]{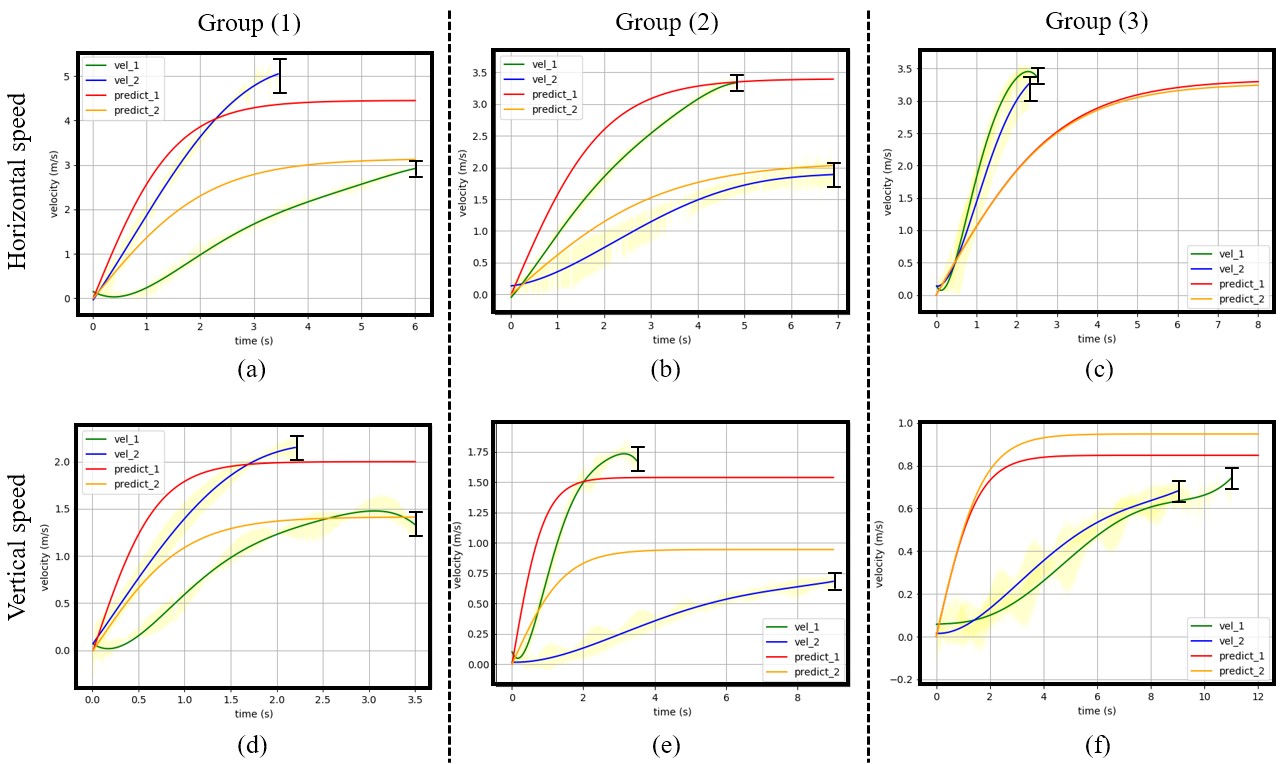}
        \caption{The performance comparison between blimps with different configuration.The first line figures are the Speed comparison in the horizontal direction and the second line figures are speed comparison in the vertical direction. The yellow areas are the error bound of 9 trials.}
    \label{exp_td}
\end{figure*}

\begin{table}[h]
    \caption{Comparison of Measured and Calculated Balloon Volumes}
    \label{table: comp}
    \centering
    \resizebox{0.9\linewidth}{!}{
    \begin{tabular}{||c||c|c||c|c|c||}
    \hline
        Balloon       & Actual Vol. ($m^3$) & Calc. Vol. ($m^3$) & $\lvert$\% Error$\rvert$ \\
    \hline
     White &          0.0338664 &          0.0326159 &      3.834 \\
        Silver        &          0.0139214 &          0.0152378 &      8.639 \\
        Red           &          0.0963472 &          0.1261541 &     23.627 \\
        Black         &          0.0721492 &          0.0947815 &     23.878 \\
       
    \hline
    \end{tabular}}

\end{table}

In this case study, we conducted measurement and comparison on 2 saucer-shaped balloons (Silver and Red), 1 oval shape balloon (Black), and 1 spherical shape balloon(white) in order to verify the theory from payload estimation. We measure their actual volume after inflated and we also calculate their theoretical volume using our theory. As shown in Table \ref{table: comp}, all the mean error is less than $25\%$, which means that our calculation can be regarded as the upper bound of the payload estimation.  

\subsection{Experiment settings}

We considered 3 groups of experiments to verify the impact of different design parameters. In order to verify the design evaluation system, we conduct performance testing under the Opti-track system and compared it with the results from the evaluation system.

In the 3 groups of experiments, We use the controlled variable method to change only one of the parameters and keep the other parameters unchanged each time as shown in Table \ref{table:settings}. In experiment group (1), we use different thrust distributions and maintain all other high-level parameters as the same. In group (2), we choose two different propulsion combinations. In experiment group (3), we used two different balloons that have different shapes. 

\begin{table}[]
\caption{The Experiment settings}
\label{table:settings}
\begin{tabular}{c|ccc}
\hline
\begin{tabular}[c]{@{}c@{}}Design\\ parameters\end{tabular} & \begin{tabular}[c]{@{}c@{}}Thrust\\ Distribution\end{tabular} & \begin{tabular}[c]{@{}c@{}}Propulsion\\ combination\end{tabular}                                              & \begin{tabular}[c]{@{}c@{}}Balloon\\ Shape\end{tabular} \\ \hline
\multirow{2}{*}{Group (1)}                                  & Table I case 1                                                & \multirow{2}{*}{\begin{tabular}[c]{@{}c@{}}d\_p = 54mm,d\_m = 8.5mm\\ f\_t = {[}-15,15{]}g\end{tabular}}  & \multirow{2}{*}{Oval}                                   \\ \cline{2-2}
                                                            & Table I case 2                                                &                                                                                                               &                                                         \\ \hline
\multirow{2}{*}{Group (2)}                                  & \multirow{2}{*}{Table I case 2}                               & \begin{tabular}[c]{@{}c@{}}d\_p = 65mm, d\_m = 8.5mm \\ f\_t = {[}-17.5,17{]}g\end{tabular}                 & \multirow{2}{*}{Oval}                                   \\ \cline{3-3}
                                                            &                                                               & \begin{tabular}[c]{@{}c@{}}d\_p = 31mm, d\_m = 7mm\\ f\_t = {[}-3.7,6.6{]}g\end{tabular}                      &                                                         \\ \hline
\multirow{2}{*}{Group (3)}                                  & \multirow{2}{*}{fig. 2 case}                                  & \multirow{2}{*}{\begin{tabular}[c]{@{}c@{}}d\_p = 46.5mm, d\_m = 7mm\\ f\_t = {[}-5.6,2.8{]}g\end{tabular}} & Saucer                                                  \\ \cline{4-4} 
                                                            &                                                               &                                                                                                               & Oval                                                    \\ \hline
\end{tabular}
\end{table}

\subsection{Results and discussion}
According to the comparison results, as shown in Fig. \ref{exp_td}, the evaluation system is able to predict the vertical and horizontal speed trend, and it can also predict the terminal velocity of the designed airship. Some of these predictions are even within the error range of real experiments, for example, the horizontal speed of group (2)(b) and group (3)(c), and the vertical speed of group (1)(d). Except for the Fig. \ref{exp_td} (f), the maximum prediction errors is less then $25\%$. Based on the experiment data from Fig. \ref{exp_td} (f), we have reason to believe that in the real experiments, the actually designed blimp didn't reach its maximum velocity before the Opti-track system stopped recording. 
Some other observations from the experiments:
\begin{itemize}
    \item In the group (1) experiments, although the thrust in the first configuration is twice the thrust in the second configuration, since the drag coefficient increases with the increase in speed, the terminal speed is not twice the relationship. 
    \item Even though the thrust distribution is the same, different motor + propellers combinations will also cause a huge impact on the performance of the designed blimp. And change the propulsion system is actually the fastest way to change its performance. 
    \item Balloons with different shapes may "accidentally" have similar performance, but due to different volumes, the payload could be very different. 
\end{itemize} 

\section{CONCLUSIONS}
This work provides a modular and reconfigurable design framework to help the non-expert user design their own custom blimps. By exploring the design parameter space, our design and performance evaluation system provides an end-to-end solution that will both build and characterize the blimp designed by users. Using a case study and comparison between experiments and predicted performance, we have proved the feasibility of the design and evaluation system presented in this paper.

\addtolength{\textheight}{-12cm}   % This command serves to balance the column lengths
                                  % on the last page of the document manually. It shortens
                                  % the textheight of the last page by a suitable amount.
                                  % This command does not take effect until the next page
                                  % so it should come on the page before the last. Make
                                  % sure that you do not shorten the textheight too much.

%%%%%%%%%%%%%%%%%%%%%%%%%%%%%%%%%%%%%%%%%%%%%%%%%%%%%%%%%%%%%%%%%%%%%%%%%%%%%%%%

%%%%%%%%%%%%%%%%%%%%%%%%%%%%%%%%%%%%%%%%%%%%%%%%%%%%%%%%%%%%%%%%%%%%%%%%%%%%%%%%

%%%%%%%%%%%%%%%%%%%%%%%%%%%%%%%%%%%%%%%%%%%%%%%%%%%%%%%%%%%%%%%%%%%%%%%%%%%%%%%%
% \section*{APPENDIX}

% Appendixes should appear before the acknowledgment.

% \section*{ACKNOWLEDGMENT}

% This paper is supported by Q.? The author would like to thank his labmate Wenzhong for helping him understand the knowledge of batteries. The author would also like to thank Yifei Chen, Ryan Chen, Pranav Sankar Srinivasan for their help on developing the project relative technique.  

%%%%%%%%%%%%%%%%%%%%%%%%%%%%%%%%%%%%%%%%%%%%%%%%%%%%%%%%%%%%%%%%%%%%%%%%%%%%%%%%

\bibliography{references}{}
\bibliographystyle{unsrt}

\end{document}